# Electrochemical kinetics and dimensional considerations at the nanoscale

H. Yamada[1*], and P.R. Bandaru[1, 2*]

[1] Department of Electrical Engineering,

[2] Program in Materials Science, Department of Mechanical Engineering,

University of California, San Diego, La Jolla, CA

**Abstract**

It is shown that the consideration of the density of states variation in nanoscale electrochemical systems yields modulations in the rate constant and concomitant electrical currents. The proposed models extend the utility of Marcus-Hush-Chidsey (MHC) kinetics to a larger class of materials and could be used as a test of dimensional character. The implications of the study are of much significance to an understanding and modulation of charge transfer nanostructured electrodes.

[*] E-mail: pbandaru@ucsd.edu

A critical understanding of the thermodynamics and kinetics inherent to electrochemical reactions is necessary for scientific insights into charge transfer [1] as well as in applications ranging from biochemical reactions [2] to charge storage in capacitors and batteries [3]. While the foundational attributes have almost always been reckoned in terms of one-electron based charge transfer [4,5], much of the theoretical and experimental analysis has only obliquely referred to the considerations of dimensionality. Consequently, three-dimensional electrode characteristics and classical thermodynamics have been implicitly assumed in heterogeneous electron transfer kinetics, encompassing the widely used Butler-Volmer (BV) formulations and the subsequent Marcus [6,7] – Hush [8] interpretations. In this regard, Arrhenius based activation theory, leading to the BV approaches, has been used for over a century, and extensively documented in standard electrochemistry textbooks [4,9]. In the BV case, the rate constant ($K^{BV}$), considering that for the forward reaction rate ($K_F$) and for the backward reaction ($K_B$), is:

$$K^{BV} = K_F + K_B = K^o \exp\left[\frac{\alpha e \eta}{k_B T}\right] + K^o \exp\left[\frac{(1-\alpha) e \eta}{k_B T}\right] \tag{1}$$

In Eqn. (1), the $\alpha$ is the electron transfer coefficient and $\eta$ refers to the overpotential (= $V - V^o$), with $V$ as the applied voltage and $V^o$ as the standard redox potential. The $e$ is the elementary unit of electronic charge, $k_B$ is the Boltzmann constant, and $T$ is the temperature. While simple to use, in principle, such an approach does not yield substantial insight into the type and involvement of the specific constituents (redox species as well as the electrode) and the $\alpha$ is phenomenologically determined. The Marcus-Hush (MH) theory then seeks to better understand the rationale for the $K^o$ and $\alpha$ through a more detailed consideration [10] of the reorganization dynamics of the solvent and the redox species *vis-à-vis* the electrochemical reactions and the electrolyte (through the macroscopic dielectric constant).

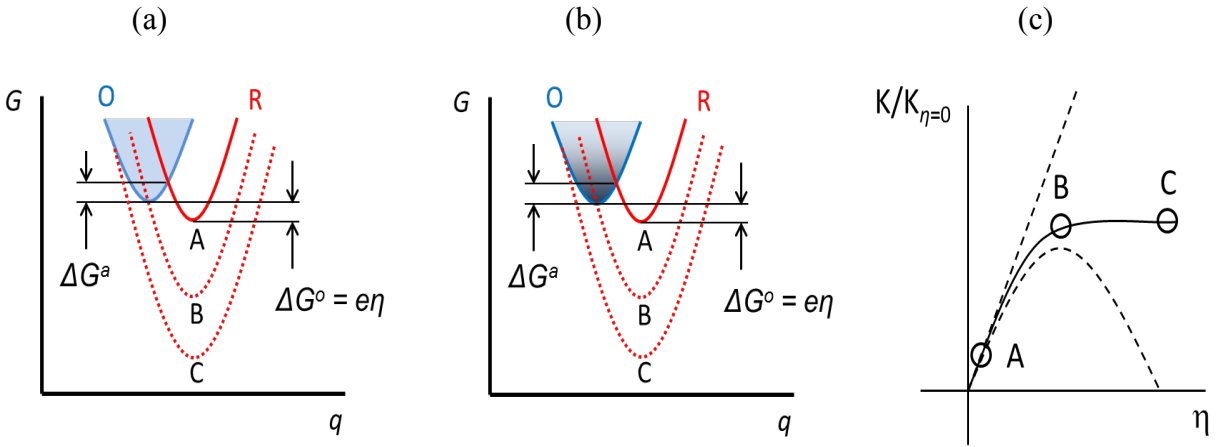

**Fig. 1(a)** Gibbs free energy ($G$) - reaction coordinate ($q$) curves for the oxidized ($O$) and reduced ($R$) species in an electrochemical redox reaction, of the type: $O + e^- \rightleftharpoons R$. The free energy of reaction ($\Delta G^o$), and the free energy of activation ($\Delta G^a$) are indicated, for several applied voltages ($\eta$) and resulting $R$ species configurations. **(b)** For low-dimensional structures, variation in the density of states (*DOS*) accessible for electron transfer, as in the lowering of the *DOS* for a one-dimensional systems, would yield equivalent behavior, **(c)** The operating points corresponding to the various $R$ curves in (i) and (ii) represented in the normalized rate constant ($K/K_{\eta=0}$)-$\eta$ plot.

We first briefly review the salient features of the MH kinetics approach and its extension by Chidsey [11]. Consequently, we consider typical [6,7] free energy ($G$) – reaction coordinate ($q$) curves: Fig. 1. Typically, the reaction coordinate has been broadly interpreted [12], and may refer to the distance [13], in a multi-dimensional *extensive* variable sense (*e.g.*, the change of bond length, electrical charge, *etc.*), between the oxidized ($O$) and reduced ($R$) species in an electrochemical redox reaction, of the type: $O + e^- \rightleftharpoons R$. While the progressive lowering of the minimum energy of the $R$ parabola (*e.g.*, through increasing the $\eta$) always decreases the free energy of reaction $\Delta G^o$, the *free energy of activation* $\Delta G^a$ initially decreases, reaching zero when the *R* parabola passes through the minimum of the *O* parabola, and subsequently increases, due to a shift of the *R* free energy curves to the left hand side of the *O* parabola: Fig. 1(a). The concomitant increase and decrease of the electrochemical reaction rate constant $K^{MH}$, *i.e.*, as represented in Eqn. (2), with $v$ as the attempt frequency, reaches a maximum when $\Delta G^a = 0$.

$$K^{MH} = v \exp\left[-\frac{\Delta G}{k_B T}\right] = v \exp\left[-\frac{(\lambda \mp e\eta)^2}{4\lambda k_B T}\right] \quad (2)$$

Such a non-intuitive increase and subsequent decrease of the reaction rate with increasing driving force (*i.e.*, $\eta$) constitutes the essence of the *inverted region*, particular to the Marcus-Hush theory. Such a notion on the maximum of a rate constant has been experimentally confirmed [14], *e.g.*, in intramolecular reactions, concerning molecules with bridged donor – acceptor units [15]. It may also be derived that [14], $\Delta G^a = \frac{\lambda}{4}\left(1 + \frac{\Delta G^o}{\lambda}\right)^2$, with $\lambda$ as the reorganization energy - which is related to the energy required for both the internal (*e.g.*, due to the bond configuration changes) and the external (*e.g.*, in the rearrangement of the solvation shell, surrounding electrolyte, *etc.*) configurational changes. Subsequently, it is evident that a zero $\Delta G^a$ would imply that the peak of the $K^{MH}$ is at a value of $\lambda \sim - \Delta G^o$.

However, such a theory seemed to be incompatible with the notion of long distance interfacial electron transfer where the rate constant decreases exponentially with increased donor-acceptor separation distances [16] as considered through the seminal work of Chidsey [11]. Additionally, the experimental observation, in certain metal electrode based electrochemical ensembles, of the saturation of the electrochemical current with increasing $\eta$, prompted the consideration of a continuum of energy level states. The consequently derived rate constant $K^{MHC}$, considering energy level occupancy through the Fermi-Dirac distribution $f_{FD}$, and the explicit introduction of a constant metallic electronic density of states ($DOS$) (= $\rho$), was of the form [11]:

$$K^{MHC} = \nu \rho k_B T \int_{-\infty}^{\infty} \frac{1}{1+\exp(x)} \exp\left(-\frac{k_B T}{4\lambda}\left[x - \frac{\lambda \mp e\eta}{k_B T}\right]^2\right) dx \qquad (3)$$

The variable $x = \frac{E - E_F}{k_B T}$, refers to the normalized energy of a relevant participating level ($E$), *e.g.,* in the electrode, relative to the Fermi energy ($E_F$) and the negative sign is used for $\eta > 0$. The integration limits may be narrowed down to either the negative interval (-∞, 0] or the positive interval: [0, ∞), if $f_{FD}(E)$ can be approximated by a step function, which would be applicable when the $\eta$ exceeds 26 mV (=$k_B T/e$).

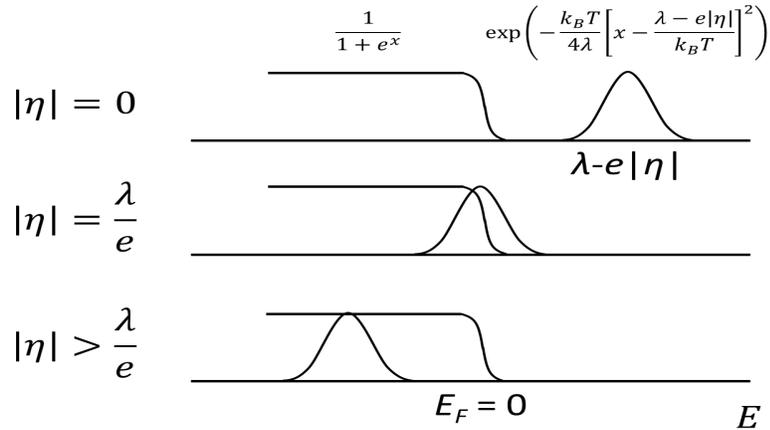

**Fig. 2** The Chidsey formulation for the rate constant in Eqn. (3) can be interpreted as indicating the relative overlap between the Fermi-Dirac distribution function $f_{FD}(E)$ – on the *left*, and the Gaussian curve – on the *right*, corresponding to the MH models.

The MHC relation, indicated in Eqn. (3), may also be interpreted as related to the area of overlap between the $f_{FD}$ and an Arrhenius based rate law: Fig. 2. This figure indicates that at low $\eta$, BV theory may be adequate to model the electrochemical kinetics while increasing $\eta$, leads to overlap and the Chidsey extension. However, at $|\eta| = \lambda/e$. and beyond, the electrical current starts to decrease and saturation may be expected when $|\eta|$ is significantly larger compared to the $\lambda/e$. The constant height of the step function results from an assumption of a constant *DOS*.

The MHC relation seems to yield excellent agreement with the experimental observation of the plateauing of the rate constant as a function of the electrode overpotential. It is to be noted that the rate constants, *e.g.,* the $K^{MHC}$ are typically obtained through chronoamperometry (CA) experiments, through the electrical current *I* decay with time *t* (in response to a step-voltage change) of the form: $I = I_o \, exp \, (-K^{MHC}t)$. However, even in such molecular systems, the Chidsey modification to the Marcus-Hush theory adopts an intrinsically continuum point of view, through assuming a constant $\rho$.

In this paper, we broadly aim to extend the utility of the Marcus-Hush-Chidsey (MHC) kinetics to a larger class of materials and situations. For instance, we observe in zero-dimensional (0-D) or one-dimensional (1-D) nanostructures, electrical current oscillations as a function of the $\eta$, corresponding to the gradual population (and de-population) of each successive sub-band. We posit that the consideration of a variable/non-constant *DOS* leads to a deeper appreciation of the MHC formulations and may yield tests of dimensional character and concomitant contribution to electrochemical systems.

First, we reinterpret the classical free energy – reaction coordinate curves depicted in Fig. 1(a), in the context of lower dimensional structures. The initial decrease in $\Delta G^a$ followed by a subsequent increase, can be related by analogy to the availability and subsequent lack in the number of energy levels (related to the *DOS*) accessible for electron transfer. Such a modulation is apparent in the *DOS* of one-dimensional nanostructures [17], with increasing carrier concentration and change of the $E_F$, and may be induced through appropriate $\eta$. We have then observed that such non-constant *DOS* yields novel electrical current – voltage response in related electrochemical systems. Fig. 1(b) indicates the correspondence for lower dimensional systems where the decreasing DOS at higher energy may be taken analogous to the increasing $\lambda$. Indeed, saturation of the electrical current/rate constant curves may be indicative of the limit of a finite *DOS*.

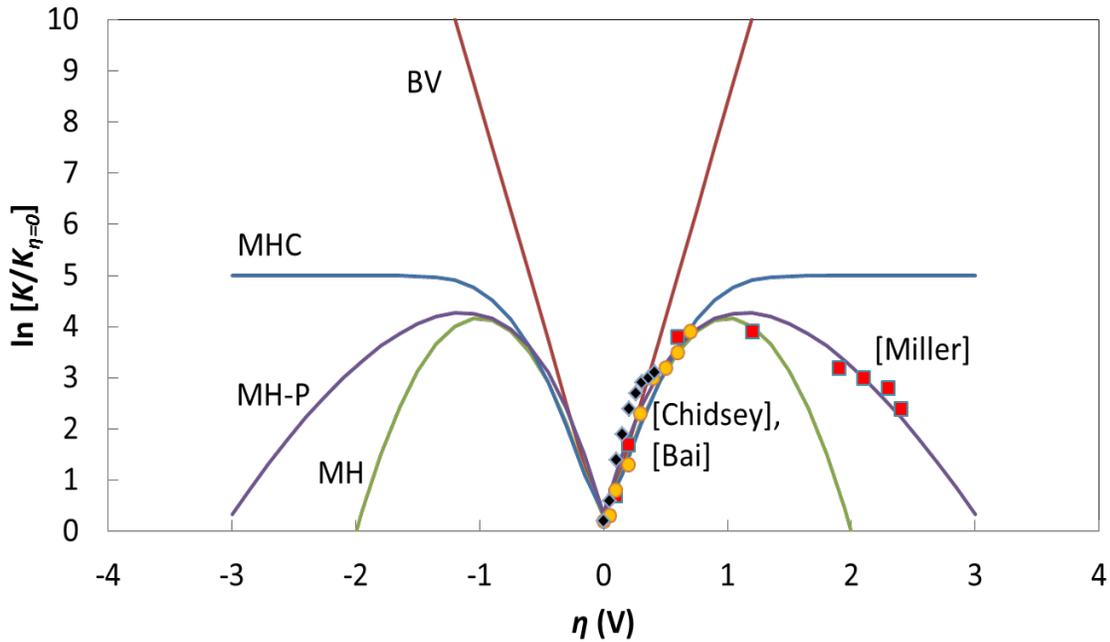

**Fig. 3** The variation of the normalized reaction rate constants, *K,* as a function of electrochemical kinetics, *i.e.,* corresponding to Butler – Volmer: BV, Marcus-Hush: MH, Poisson-Marcus: PM, or Marcus-Hush-Chidsey: MHC, models - with the overpotential $\eta$. The *K* values have been normalized to their minimum values in each case. The experimental values were adapted from Miller, *et al.* [15], Chidsey [11], and Bai, *et al,* [19].

The consequent plots of the respective normalized *K* value variation, through Eqns. (1), (2), and (3), with the $\eta$ (corresponding to BV, MH, or MHC kinetics) are indicated in Fig. 3. From CA related experiments and $I=I_o \exp(-K^{MHC}t)$, such characteristics may be considered equivalent to electrical current $I$-$\eta$ plots. The figure also indicates a re-plotting of experimental data previously obtained [15]. It is to be noted that while the BV kinetics indicates a linear variation with $\eta$, the MH model exhibits a peak as a function of the $\eta$. It is also relevant to note that the experimental curves were also fit through employing Poisson statistics [18] assuming homogeneous charge transfer, with a net $\lambda$ (=1.2 eV) constituted from (i) an *external* solvation energy $\lambda_s$ = 0.75 eV, and (ii) an *internal* vibrational energy component $\lambda_v$ = 0.45 eV. The incorporation of Poisson (*cf.,* Gaussian distribution) statistics also yields an inversion of the $K$, while avoiding the steeper drop-off of the MH curve, and was considered [15] the best fit to certain chronoamperometric data.

We now consider the influence of a variable *DOS*, on the $K$ variation with $\eta$. The number of electrons available for the redox reaction from the electrode: $n$, would be: $n = \int_{E_C}^{\infty} f_{FD}(E - E_F) DOS(E - E_c) dE$, where $E_c$ is the energy at the bottom of the conduction band. We concomitantly introduce a new *DOS* based reaction rate constant: $K^{MHC-DOS}$, considering the influence of the energy levels, through:

$$K^{MHC-DOS} = \nu k_B T \int_{-\infty}^{\infty} \frac{DOS\left(\left|x + \frac{E_F - E_c}{k_B T}\right|\right)}{1 + \exp(x)} \exp\left(-\frac{k_B T}{4\lambda}\left[x - \frac{\lambda \mp e\eta}{k_B T}\right]^2\right) dx \quad (4)$$

The integration may again be either over the negative interval (- $\infty$, 0] or the positive interval: [0, $\infty$), as previously discussed. In a limiting case corresponding to Eqn. (3), the *DOS* would be a constant (*e.g.,* $\rho$), reverting to the original Chidsey formulation [11]. In the subsequent treatment, the $E_c$ was taken as reference energy and set to zero. Such a formulation involving the energy variation of the *DOS* [17] as a function of the dimensionality, $D$ (*e.g., $DOS_{3D}$* ~ constant or ~

$E^{1/2}$ – for a semiconductor, $DOS_{2D} \sim E^0$, $DOS_{1D} \sim E^{-1/2}$, $DOS_{0D} \sim$ Dirac delta function like) also allows for a variable height of the step function, depicted on the left hand side of Fig. 2. The resulting $K^{MHC\text{-}DOS}$-$\eta$ curves, as a function of the dimensionality dependent *DOS* are indicated in Fig. 4. In addition to the parabolic *Energy-k vector* dispersion, we have also incorporated a linear *E-k* dispersion as seems to be necessary to describe the characteristics of graphene and related 2D materials [20,21].

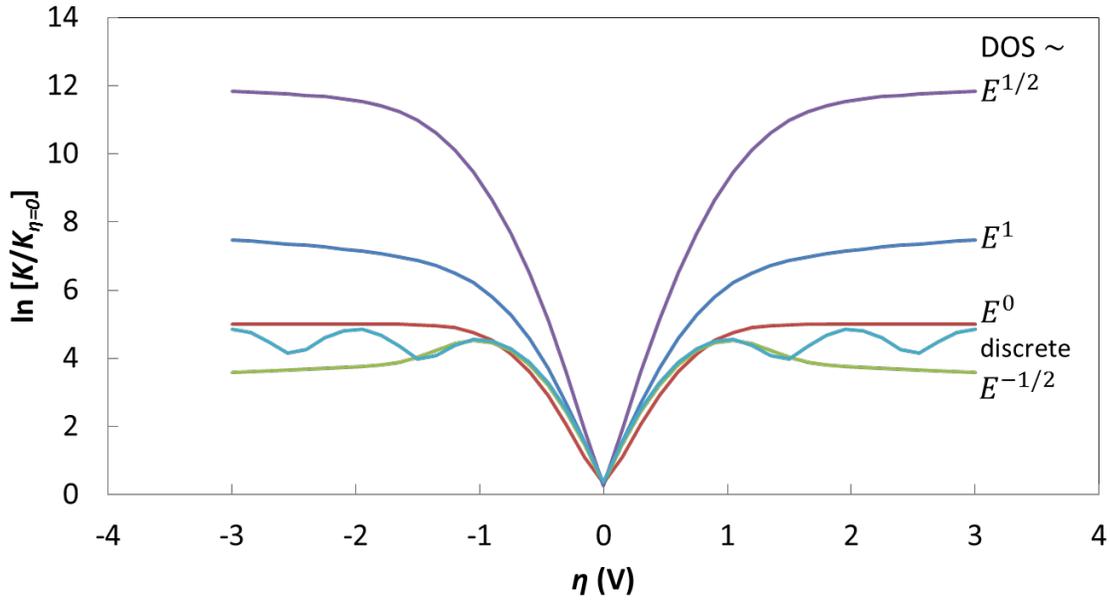

**Fig. 4** The variation of the normalized $K^{MHC\text{-}DOS}$ based reaction rate constants with $\eta$ - obtained from Eqn. (4), for electrodes with $DOS \sim E^a$. The exponent *a*, indicated in the figure, is a function of the electrode dimensionality (*i.e.*, $a = ½$ for a three-dimensional semiconductor; $a = 0$ or 1, for a two-dimensional system, $a = -½$ for a one-dimensional system) and is Delta-function like for zero-dimensional systems, such as quantum dots. The case of $a = ½$ involves a bandgap, which causes the $K_{\eta=0}$ to be smaller than that for the other cases. Generally, a reduction of the *K* corresponds to a decreasing DOS with energy.

The respective influences of the dimensionality and the dispersion are clearly evident. While the traditional MHC based formulations assumed a constant *DOS*, particular to bulk-like/three-dimensional (3D) *metallic* electrodes, the energy variation of the *DOS* in lower dimensional systems yields rich and involved behavior. For instance, the behavior of a two-dimensional (2D) material with parabolic energy dispersion, *e.g.,* involving a quantum well, is seen to differ compared to one with linear energy dispersion, *e.g.,* graphene. In the latter case, an increasing *DOS* with electron kinetic energy is responsible for the observed variation. The situation for a one-dimensional (1D) material, *e.g.,* a carbon nanotube (CNT), constituted electrode - with parabolic energy dispersion along the long-axis and quantization along the two perpendicular directions, with a decreasing *DOS vis-à-vis* energy, corresponds to inversion in the $K$-$\eta$ curves at a sufficiently large $\eta$, as posited in the original MH formulations. In one-dimensional systems, the initial increase of the DOS upon the $E_F$ reaching the band edge and the subsequent $E^{-1/2}$ induced decrease yields a corresponding modulation of the $K$ and the electrical currents.

We then predict the occurrence of oscillations in the $K/K_{\eta=0}$ - $\eta$ curves in one-dimensional nanostructures as a function of chirality in Fig. 5. As is well known [22], the specific nature of wrapping of a constituent graphene sheet, through the chirality index [*m, n*], dictates whether the resulting CNT is metallic/semiconducting. We depict the corresponding *DOS* for a (i) semiconducting [10,0] nanotube, and a (ii) metallic [9,0] nanotube: Figures 5(b) and (c), respectively. While the rate constant oscillations are particularly pronounced in the former, they are less so in the latter case. The underlying reason may be related to the smaller (/larger) separation of the energy sub-bands, respectively. Moreover, the oscillations in the semiconducting and the metallic cases occur at different voltages, corresponding to the *DOS* variation.

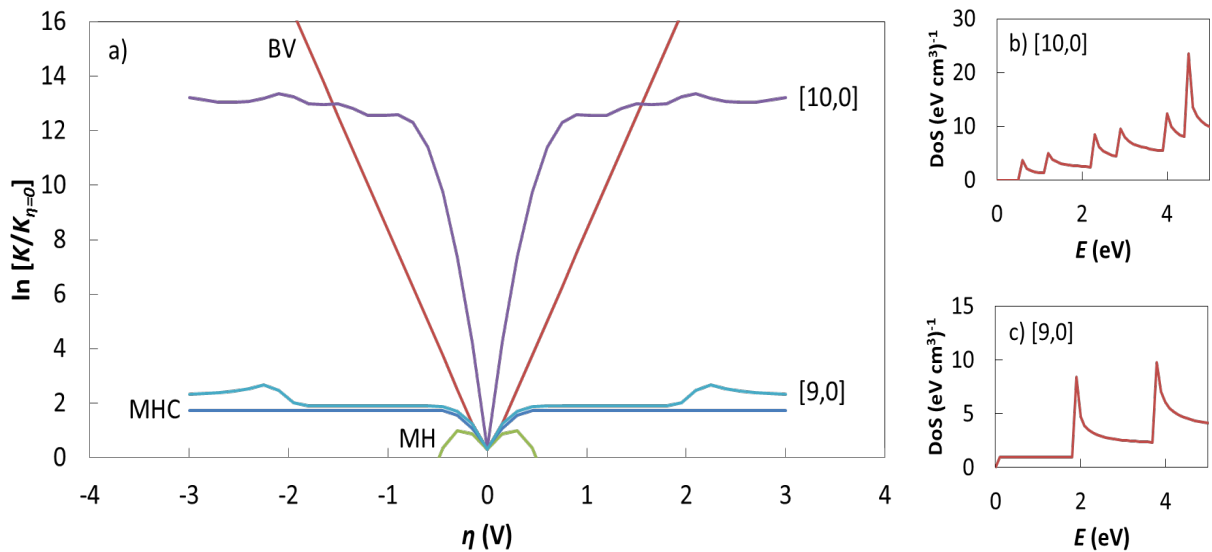

**Fig. 5 (a)** The predicted variation of the reaction rate constants with $\eta$ for a [10,0] and [9,0] single-wall CNT with $\lambda$ = 0.25 eV. The respective *DOS* (*E*) variation, for the **(b)** [10,0], and **(c)** [9,0] single-wall CNT are indicated. The modulations in the rate constants are due to the successive population and de-population of sub-bands in the nanostructure. The $E_F$ was taken to be zero in these simulations.

We also noted that the width of the energy sub-bands ($\Delta E$) in the electrode considered on the horizontal axis of *DOS* (*E*), with respect to $\lambda$ - typically electrode adjacent (*e.g.,* electrolyte) side, would be another important variable in interpretation of the *K-$\eta$* or the resultant *I-$\eta$* modulations. Generally, $\Delta E$ is indicative of the energy level spacing and inversely related to the size of the nanoscale electrode, *e.g.,* in a one-dimensional CNT of diameter *d*, the equivalent $\Delta E$ would be proportional to $1/d^2$. The magnitude of the $\lambda$, as necessary to proceed from an *O* states to an *R* state (as in $O + e^- \rightleftharpoons R$), *cf.,* Fig. 1, can be considered analogous to an energy level width. When the $\Delta E$ is larger (/smaller) compared to the $\lambda$, the interaction of the electrode energy levels (and relevant electron exchange/redox interactions) with respect to the electrolyte would be more (/less) sharply defined, and yield an oscillatory (/smooth) *K-$\eta$* variation. A small $\lambda$ implies that the nuclear reconfiguration and the coordinating solvent interactions [23] accompanying the redox reaction is negligible. At a large enough $\lambda/\Delta E$, a continuous electronic distribution/*DOS* may be assumed, yielding smooth MHC kinetics, with an increase of the *K* up to $\eta \sim \lambda/e$, and subsequent plateauing of the *K-η* curves. The discussed *K-η* variation as related to the $\lambda/\Delta E$ ratio is indicated in Fig. 6.

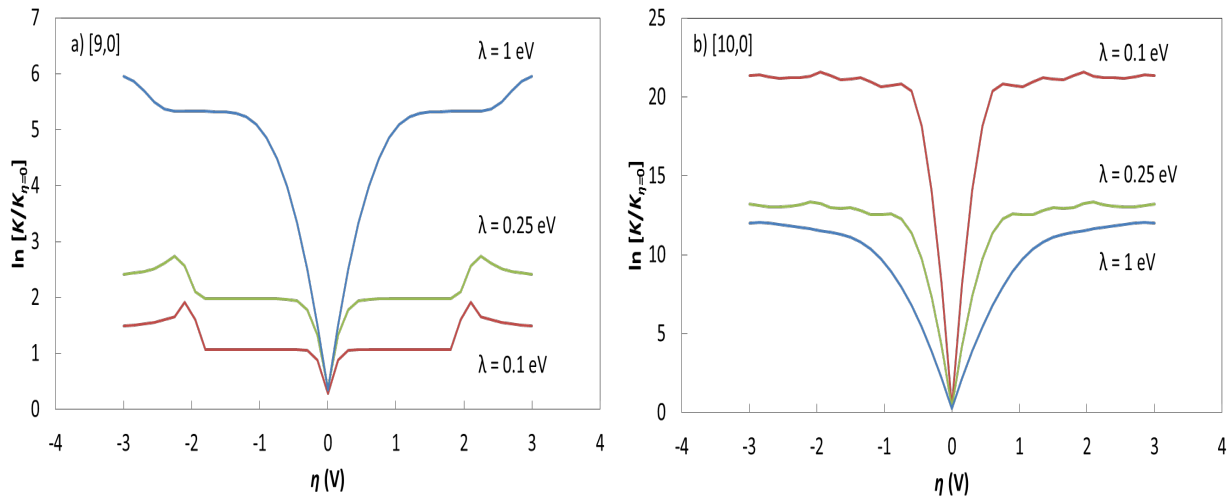

**Fig. 6** The $K/K_{\eta=0}$ variations for a **(a)** [9,0] CNT, and **(b)** [10,0] CNT, as a function of the $\lambda$, indicates the competing effects of the classical reorganization energy ($\lambda$) and the sub-band energy separation ($\Delta E$), with respect to the influence of the *DOS*. The bandgap for the [10,0] CNT actually causes $K/K_{\eta=0}$ to decrease with increasing $\lambda$, in contrast to the [9,0] CNT.

As it was recently indicated that a $\lambda$ of ~ 0.2 eV seemed to be effective for modeling MHC based charge transfer kinetics at LiFePO$_4$ battery electrode interfaces [19], such modulations could be experimentally probed. Additionally, the $K/K_{\eta=0}$ increases with $\lambda$ for a [9,0] CNT, as was previously indicated [10], but shows the opposite variation in a [10,0] CNT. The bandgap in the semiconducting [10,0] CNT causes the $K_{\eta=0}$ value to be smaller than that for the metallic [9,0] CNT; such an effects is stronger for smaller $\lambda$, *cf.* Fig. 2.

In summary, we have shown that considering the specific nature of the *DOS*, as would be necessary in nanostructured materials, leads to a modification of the expected MHC electro-kinetics. We have predicted, most notably, the occurrence of oscillations of the rate constant and the concomitant electrical current in semiconducting nanotubes, the experimental verification of which would be a significant test of the nature of electrical conductivity as well as dimensionality. The implications of our study would be relevant to the use of nanostructured electrodes in electrochemical storage systems where such electrical current modulations would impact energy and power delivery.

The authors are grateful for support from the Defense Advanced Research Projects Agency (DARPA: W911NF-15-2-0122), National Science Foundation (NSF: CMMI 1246800), and acknowledge detailed discussions and interactions with R. Narayanan and Prof. P. Asbeck.